# Local disorder in $Na_{0.5}Bi_{0.5}TiO_3$-piezoceramic determined by 3D electron diffuse scattering

Alexandra Neagu[1] and Cheuk-Wai Tai[1]

Department of Materials and Environmental Chemistry,
Stockholm University, SE-106 91 Stockholm, Sweden

## Abstract

Local structural distortions in $Na_{0.5}Bi_{0.5}TiO_3$-based solid solutions have been proved to play a crucial role in understanding and tuning their enhanced piezoelectric properties near the morphotropic phase boundary. In this work all local structural disorders in a lead-free ternary system, namely $85\%Na_{0.5}Bi_{0.5}TiO_3$-$10\%Bi_{0.5}K_{0.5}TiO_3$-$5\%BaTiO_3$, were mapped in reciprocal space by 3D electron diffraction. Furthermore, a comprehensive model of the local disorder was developed by analysing the intensity and morphology of the observed weak diffuse scattering. We found that the studied ceramics consists of plate-like in-phase oxygen octahedral nanoscale domains randomly distributed in an antiphase tilted matrix. In addition, A-site chemical short-range order of Na/Bi and polar displacements contribute to different kinds of diffuse scattering. The proposed model explains all the observed diffraction features and offers insight into the ongoing controversy over the nature of local structural distortions in $Na_{0.5}Bi_{0.5}TiO_3$-based solid solutions.

---

[1] Corresponding authors – email addresses:
Alexandra Neagu: alm.neagu@gmail.com
Cheuk-Wai Tai: cheuk-wai.tai@mmk.su.se

**Introduction**

Recent environmental regulations have questioned the use of lead-based piezoceramics in electronic applications owing to the high toxicity of lead and its derivatives, particularly lead oxide[1,2]. This has caused a recent surge in finding new lead-free piezoceramics[3,4]. One common ion used as a replacement for $Pb^{2+}$ is $Bi^{3+}$, a similarly heavy ion with a lone pair[5]. The majority of studies have focused on binary systems like $Na_{0.5}Bi_{0.5}TiO_3$-$BaTiO_3$ (NBT-BT) and $Na_{0.5}Bi_{0.5}TiO_3$-$Bi_{0.5}K_{0.5}TiO_3$ (NBT-BKT) which present improved dielectric and piezoelectric properties[6,7]. Recently, a ternary compound (NBT-BKT-BT), which has comparable dielectric and piezoelectric properties to those of lead-based piezoceramics, has been proposed. Similar to Pb-based piezoceramics, the enhanced properties of these Pb-free compounds are found at the MPB boundary[6,8-11] and are of interest for electrical and electromechanical applications.

Despite a significant number of studies there is still an ongoing debate over the average and local structures of NBT and its solid solutions. Early diffraction studies[12-14] suggest a rhombohedral $R3c$ symmetry for NBT with polar cation displacements combined with antiphase oxygen octahedral rotations ($a^-a^-a^-$ tilt system in Glazer notation[15]). More recent high-resolution X-ray studies[16,17] propose, on the other hand, a monoclinic $Cc$ symmetry with $a^-a^-c^-$ octahedral tilting. Another proposed structure suggests that the equilibrium state of NBT at room temperature consists of a coexistence of $R3c$ and $Cc$ phases[18]. The existing controversy over the average structure of NBT was further enhanced by several TEM studies that confirmed both the rhombohedral[19] and monoclinic[20] symmetries. Moreover, there is also a mismatch between the macroscopic properties and the temperature-dependent phase transitions. Structural studies propose two phase transitions. One at ~296°C from a ferroelectric rhombohedral $R3c$ structure to a ferroelectric tetragonal $P4bm$ structure and one at ~566°C from a ferroelectric tetragonal $P4bm$ structure to a paraelectric cubic $Pm\bar{3}m$ structure[14]. However, the piezoelectricity in poled NBT ceramics is lost upon heating at ~187°C[21,22], so called depolarization temperature. The difficulty in assigning a definitive average structure for NBT-based materials and the mismatch with the macroscopic properties stems from the short range order/disorder present in these types of materials. Several studies have tried to tackle this problem using EXAFS[23,24], NMR[25], X-ray diffuse scattering[26-31], TEM[32-35], pair-distribution function (PDF)[5,36] or DFT calculations[37-41]. EXAFS measurements[23] suggest a highly distorted local coordination for Bi with a shorter minimum Bi-O bond of 2.2 Å than the one proposed by diffraction experiments, which is 2.5 Å. RMC refinement of PDF data reveals an anisotropic Bi-O bonding[5,36] which suggest that Bi is strongly displaced off-center in the oxygen polyhedron. Most X-ray diffuse scattering studies[26-28,31] report two types of diffuse features: (i) broad diffuse regions around all Bragg peaks that have been modelled using a pseudorandom occupation of the A-site by Na/Bi together with an atomic size effect[26] and (ii) "L-shaped" diffuse streaks that emanate from the Bragg peaks and extend towards lower q-values which have been explained by assuming the presence of nanometer-scale platelet structures of highly correlated Na/Bi atomic positions, analogous to Guinier Preston zones (GPZs)[26]. From TEM studies two models have been proposed in order to explain the observed superstructure reflections and diffuse scattering intensity. One consists of tetragonal platelets with $a^0a^0c^+$ oxygen octahedral tilt distributed in a rhombohedral matrix with $a^-a^-a^-$ octahedral tilt[33,34] while the second one consists of pseudo-rhombohedral assemblages of nanoscale orthorhombic domains that present $a^-a^-c^+$ octahedral tilting[35].

Even though NBT and its solid solutions with BT and BKT have been intensively studied a deep understanding of the local structure in NBT-based compounds is still lacking. In this work a novel rotation electron diffraction (RED) method[42,43] was used to investigate the long- and short-range structure for a NBT-ternary ceramic, namely $85\%Na_{0.5}Bi_{0.5}TiO_3-10\%Bi_{0.5}K_{0.5}TiO_3-5\%BaTiO_3$ (85NBT-10BKT-5BT) with optimized composition near the MPB[9-11,44,45]. The RED method has already been applied with success for determination of the average structure in complex materials such as zeolites[46]. The use of electrons instead of X-rays or neutrons for diffuse scattering studies has a significant advantage since electrons interact stronger with matter ($10^6$ stronger than X-rays). The strong electron-matter interaction makes it possible to easily record superstructure reflections and weak diffuse scattering intensity with good signal-to-noise ratio. Also the fact that the RED method provides a 3D reciprocal-space volume makes the analysis of diffuse scattering intensity more straight forward than in the case of 2D electron diffraction patterns. Furthermore DISCUS[47,48] software was used to generate disordered atomic structures and compute the corresponding calculated data, i.e. electron diffraction patterns. Here, we propose a structural model to explain all the observed electron diffuse scattering intensity and superstructure reflections. To our knowledge this is the first time 3D electron diffuse scattering result together with a comprehensive structural model for understanding the local disorder in a NBT-ternary compound are reported.

**Results**

**3D electron diffuse scattering.** Due to the strong electron-mater interaction the RED method is well suited for studying small local structural deviations. Our approach is to use this technique to finely map the reciprocal-space for materials with disordered structures and record simultaneously both Bragg reflections and diffuse scattering intensity. Earlier studies of both pure NBT[26,28,33,35] and NBT-BT[29-31] have revealed a highly disordered structure for these materials, so 85NBT-10BKT-5BT ternary solid solution represents a good candidate for the study of short-range order/disorder by 3D electron diffraction. Figure 1a displays a TEM image of the grain used for RED data collection. Care was taken to avoid the damaged amorphous edge, so the SAED aperture was placed in the middle of the grain during the entire data collection. A general view of the reconstructed 3D reciprocal-space volume displaying both Bragg reflections and diffuse scattering intensity is shown in Fig. 1b. Since the reconstruction of the 3D reciprocal-space volume implies simply combining the individual 2D SAED patterns, the reconstructed result does not suffer from distortions produced by the missing wedge as is the case for tomography where back-projection algorithm is used. In Fig. 1(c-e) a schematic representation of the 3D reciprocal-space is shown in order to differentiate between a slice/volume-section through the 3D reciprocal-space volume and the whole reciprocal-space volume. A slice along a certain plane is equivalent to a typical SAED pattern. A big advantage of the RED method is that one can cut slices that are not at zero Laue-zone which simplifies the analysis of diffuse scattering intensity. A thin volume-section, centered along a particular slice, was generally used to highlight certain diffuse scattering features that were not clearly visible in the single slice. The lack of visibility of diffuse scattering intensity at low order ZAs is discussed later in the section. On the other hand, viewing the whole reciprocal-space volume oriented along different directions enabled a 3D analysis of the diffuse scattering features.

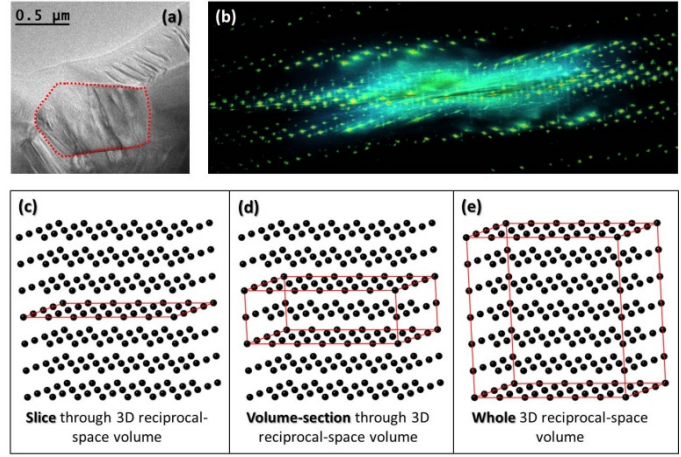

**Figure 1. 3D reciprocal-space volume. (a)** TEM image of the 85NBT-10BKT-5BT grain used for RED data collection i.e. the area delimited by the red contour. **(b)** General view of the reconstructed 3D reciprocal-space volume displaying both Bragg reflections and diffuse scattering intensity. **(c)** Schematic representation of a slice through the 3D reciprocal-space volume. **(d)** Schematic representation of a volume-section through the 3D reciprocal-space volume. **(e)** Schematic representation of the whole reciprocal-space volume.

In Fig. 2(a-c) the whole 3D reciprocal-space volume for 85NBT-10BKT-5BT ternary ceramic was projected along the three main pseudocubic directions $[001]_{pc}$, $[010]_{pc}$ and $[100]_{pc}$, respectively. Besides the fundamental perovskite reflections three extra features have been observed. The first feature comprises of two types of superstructure reflections ½(*ooe*) and ½(*ooo*) (where *o* stands for odd and *e* for even). All three variants of the ½(*ooe*) superstructure reflection, namely ½(*ooe*) ½(*oeo*) and ½(*eoo*), are present as confirmed by the fact that they can be seen along all three main pseudocubic directions (Fig. 2(a-c)). Moreover all three variants can be simultaneously observed in the $[111]_{pc}$ zone axis (supplementary Fig. 1c). The ½(*ooo*) superstructure reflections were observed for $[011]_{pc}$ zone axis as shown in supplementary Fig. 1b. The second feature is broad diffuse scattering intensity near the fundamental perovskite reflections. This feature can be clearly seen for reflections close to the direct beam i.e. 110, -110, 1-10, -1-10 (Fig. 2a) and tends to be concentrated at positions in between fundamental perovskite reflections. The third feature comprises of continuous diffuse scattering rods (Fig. 2(a-c)) along all 3 main pseudocubic directions $[00l]_{pc}^*$, $[0k0]_{pc}^*$ and $[h00]_{pc}^*$ respectively. For simplicity and convenience of comparison pseudocubic axes ($Pm\bar{3}m$) have been used for indexing throughout the paper.

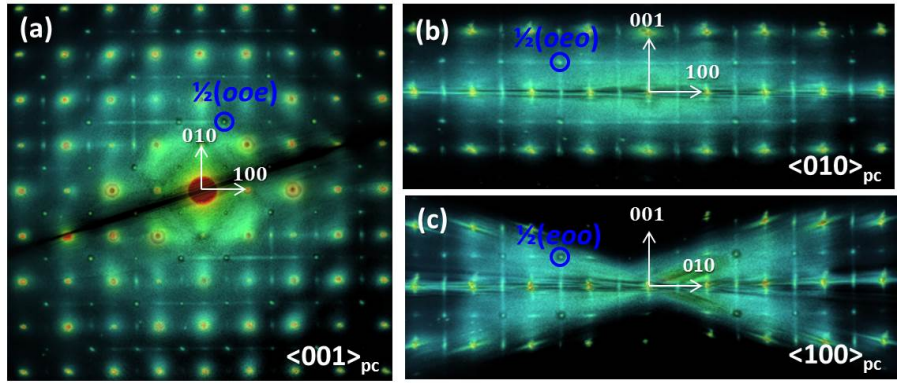

**Figure 2. 3D reciprocal-space volume projected along the 3 main perovskite pseudocubic directions.** The whole reciprocal-space volume projected along **(a)** $[001]_{pc}$ direction, **(b)** $[010]_{pc}$ direction and **(c)** $[100]_{pc}$ direction, respectively.

Figure 3 shows the projection of the whole reciprocal-space volume along $[100]_{pc}$ direction together with a volume-section centered on the *hk0* plane and the *hk0.5* slice. A closer look reveals the fact that the diffuse scattering rods are present only in *hk0.5*-type slices, as can be seen in Fig. 3c. Moreover, increased intensity can be observed at positions of *½(ooe)* reflections as indicated by white arrows. In order to highlight the weak broad diffuse intensity near the fundamental perovskite reflections instead of a slice, a volume-section was cut around the *hk0* plane (Fig. 3b). The projection centered on the *hk0* slice does not contain any diffuse scattering rods but broad diffuse intensity near the fundamental perovskite reflections can be clearly observed (i.e. the region centered on the direct beam depicted by the white square). A closer look at the individual electron diffraction frames reveals the fact that broad diffuse intensity can be easily observed for SAED patterns slightly away from exact ZA while close to ZA this feature seems to disappear (Supplementary Fig. 2). The lack of visibility for diffuse scattering intensity at low order ZAs has been reported before for a number of different materials[49-51]. One proposed explanation is that at exact ZA compositional and/or displacive disorder is averaged out along the atomic columns leading to no net modulation[49,50]. Another explanation is that for crystals ≥ 200 nm thick a strong channeling effect at low order ZAs leads to a smearing into the background of any diffuse scattering intensity[51].

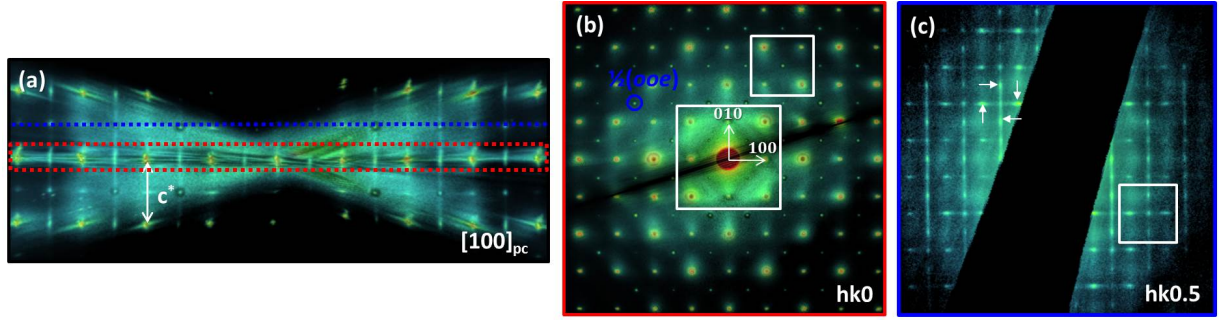

**Figure 3. Separating different diffuse scattering features using slices/volume-sections through the 3D reciprocal-space volume. (a)** The whole 3D reciprocal-space volume projected along $[100]_{pc}$. **(b)** The volume-section centered on the *hk0* slice, outlined by the red dash rectangle in (a), reveals ½(*ooe*) superstructure reflections and broad diffuse scattering intensity around the fundamental perovskite reflections. **(c)** *hk0.5* slice outlined by the blue dotted line in (a). Continuous diffuse scattering rods can be observed along $[h00]_{pc}^{*}$ and $[0k0]_{pc}^{*}$ directions with increased intensity at positions of ½(*ooe*) reflections as indicated by white arrows.

The 85NBT-10BKT-5BT ternary compound has a composition near the MPB[9-11,44,45] where a competition between rhombohedral (*R3c*) and tetragonal (*P4bm*) symmetries exists. The rhombohedral symmetry is characterized by $a^-a^-a^-$ antiphase tilt system which produces ½(*ooo*) superstructure reflections while the tetragonal symmetry is characterized by $a^0a^0c^+$ in-phase tilt system which produces ½(*ooe*) superstructure reflections. Similar superstructure reflections have also been reported for NBT-BT binary compounds[52,53] and other NBT-BKT-BT compositions[11,54] near the MPB. Besides the superstructure reflections that can help us identify the type of octahedral tilting, the RED data contains diffuse scattering intensities that give insight into the short-range structure. Few synchrotron X-ray studies have reported similar diffuse scattering features for NBT[26,28] and NBT-BT single crystals[29-31]. The observed diffuse scattering intensities have been attributed to planar defects of highly correlated Na/Bi atomic positions analogous to GPZs[26,29]. However, no definitive structural model that explains both superstructure reflections and diffuse scattering intensities has been reported for this type of materials. In order to explain the observed superstructure reflections and diffuse scattering intensities three different mechanisms of structural distortion have been taken into consideration: (1) tilting of oxygen octahedra; (2) polar displacements of all ions; (3) short range chemical order on the A-site.

**Dark-field imaging of antiphase and in-phase nanosized domains.** In order to determine the size and distribution of $a^-a^-a^-$ and $a^0a^0c^+$ domains, dark-field images taken using ½($ooo$) and ½($ooe$) superstructure reflections have been recorded. Figure 4b is a representative dark-field image of $a^-a^-a^-$ domains (bright contrast), which was recorded using the *-3/2 3/2 1/2* reflection along [110]$_{pc}$ ZA as highlighted by the red circle in Fig. 4a. A non-uniform distribution can be observed for the antiphase domains with regions of large and low concentration of bright speckles, indicating that some areas are highly populated by $a^-a^-a^-$ domains (red circle in Fig. 4b) while in others few $a^-a^-a^-$ domains are found. Figure 4d is a representative dark-field image of $a^0a^0c^+$ domains. The image was recorded using the *-3/2 1/2 1* reflection along [111]$_{pc}$ ZA, as highlighted by the blue circle in Fig. 4c. In contrast to the $a^-a^-a^-$ domains a uniform distribution of the $a^0a^0c^+$ domains could be observed. The size of the $a^-a^-a^-$ domains was less than ~8 nm across while the size of $a^0a^0c^+$ domains was less than ~12 nm across which is in good agreement with other reported values[55].

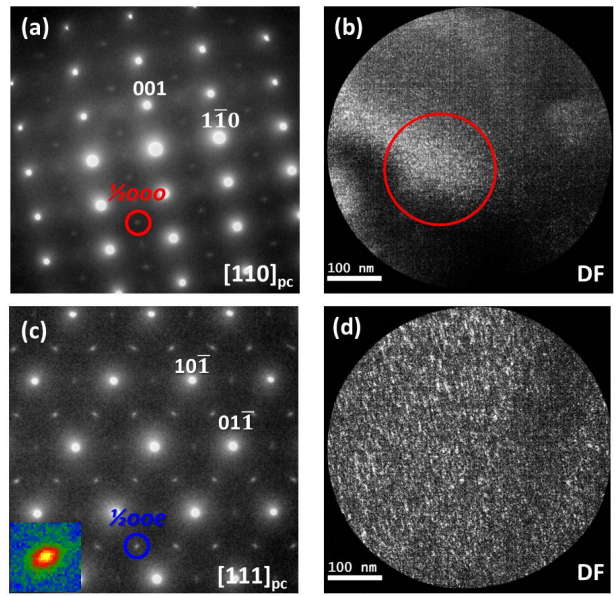

**Figure 4. Dark-field images of nanosized antiphase and in-phase octahedral tilted domains.** SAED patterns taken along **(a)** [110]$_{pc}$ and **(c)** [111]$_{pc}$ zone axes, respectively. Dark-field images of these nanosized domains taken by **(b)** ½($ooo$) superstructure reflection, as indicated by the green circle in (a) and **(d)** ½($ooe$) superstructure reflection, as indicated by the red circle in (c). The white circle in (b) highlights a region with a high concentration of $a^-a^-a^-$ domains. The inset in (c) shows an enlarged view of the slightly elongated ½($ooe$) superstructure reflection, highlighted by the red circle.

The [111]$_{pc}$ ZA simultaneously presents ½(*ooe*) ½(*oeo*) and ½(*eoo*) superstructure reflections indicating that in the case of $a^0a^0c^+$ domains a given region consists of a mixture of three variants of in-phase oxygen octahedral tilted domains, of which tilting axes are along three orthogonal directions rather than a single domain tilted along a unique orientation. Moreover, these superstructure reflections are slightly elongated as can be seen in the inset in Fig. 4c, suggesting that the in-phase tilted domains have a plate-like shape. The dark contrast in both dark-field images corresponds to regions with octahedral tilting about a different axis than the antiphase/in-phase tilting or regions without octahedral tilting. Since the nanosized octahedral tilted domains do not follow the macroscopic polarization direction or ferroelectric domain wall, we can conclude that the octahedral tilting is confined to the nanoscale[33,54,55] and the long-range order of octahedral tilting is absent.

**Modeling of local disorder**. Computer simulations of disordered structures can aid the interpretation of diffuse scattering data and both qualitative and quantitative models can be obtained in this way[56,57]. In the present work DISCUS[47,48] was used to develop a model for the local disorder in 85NBT-10BKT-5BT ternary ceramic that explains the observed superstructure reflections and diffuse scattering intensities. The experimental data revealed the presence of two types of superstructure reflections: ½(*ooe*) and ½(*ooo*). This suggests a coexistence of two different oxygen octahedral tilting systems, one in-phase $a^0a^0c^+$ and the other one antiphase $a^-a^-a^-$. Mixed octahedral tilting systems have been excluded since ½(*oee*) superstructure reflections have not been observed for this compound either by RED or SAED. Previous XRD studies[9-11,44,45] of NBT-BKT-BT ternary compounds suggest that the studied composition is near the MPB where a competition between rhombohedral and tetragonal symmetries exists. Moreover, with the RED technique continuous diffuse scattering rods were recorded for *hk0.5*-type slices. The rods are along [00*l*]$_{pc}$*, [0*k*0]$_{pc}$* and [*h*00]$_{pc}$* directions with increased intensity at positions of ½*(ooe)* reflections. Diffuse scattering rods in reciprocal-space are commonly attributed to planar defects in real-space. Therefore, the observed diffuse scattering rods are consistent with the notion that plate-like nanoscale domains of in-phase octahedral tilting are embedded in a matrix with antiphase octahedral tilting[32-34]. The model we developed builds on this idea. In addition polar displacements and short range chemical order have also been considered.

The model of local structural disorder in 85NBT-10BKT-5BT ceramic was constructed using the following methodology. First a matrix was created comprising of antiphase tilted oxygen octahedra. Further, nanometer-sized in-phase tilted domains were randomly introduced in the matrix. Last, different configurations of A-site short range chemical order and polar displacements were tested. In the case of scattering intensities DISCUS[47,48] basically calculates the Fourier transform according to the standard formula for kinematic scattering given in equation (1).

$$F(\boldsymbol{h}) = \sum_{i=1}^{N} f_i(\boldsymbol{h}) e^{2\pi \boldsymbol{h} \boldsymbol{r}_i} \quad (1)$$

The sum is over all $N$ atoms in the supercell, where $f_i$ is the atomic form factor for atom $i$ and $\boldsymbol{r}_i$ are the fractional coordinates of the atom. The sum is calculated over all points $\boldsymbol{h}$ in reciprocal space within the plane or volume of interest. The form factors are tabulated[58] and calculated once at each $\boldsymbol{h}$ for all atoms present in the crystal. Once the structure factor $F(\boldsymbol{h})$ is calculated, the scattering intensity is simply computed as $I(\boldsymbol{h}) = F(\boldsymbol{h}) \cdot F^*(\boldsymbol{h})$.

As initial model a 3D supercell made up of 50×50×50 cubic unit cells (space group $Pm\overline{3}m$ and $\boldsymbol{a} = 3.8894$ Å) was created. In the initial model all atoms occupy the ideal positions as in the cubic perovskite structure with $Bi^{3+}$, $Na^+$, $Ba^{2+}$ and $K^+$ sharing the (0.5, 0.5, 0.5) position, $Ti^{4+}$ at the (0.0, 0.0, 0.0) position and $O^{2-}$ at (0.5, 0.0, 0.0) position. The antiphase matrix was created by tilting all oxygen octahedra according to the $a^-a^-a^-$ octahedral tilt system. If the oxygen octahedra tilting is considered as a combination of individual tilts about three Cartesian axes the final structure depends on the order in which the tilts were applied[15,59]. To avoid this problem we have considered the antiphase tilting for $a^-a^-a^-$ system as a single operation about <111>$_{pc}$ axis[59]. As required by the $a^-a^-a^-$ antiphase system alternating layers have been tilted about the <111>$_{pc}$ axis with angles of 8° and -8° respectively, as schematically depicted in Fig. 6b. The value for the antiphase angle was chosen based on reported values for the octahedron tilt angle in pure NBT[14]. To ensure an anisotropic antiphase tilting the matrix was created from small antiphase tilted domains (6×6×6 unit cells) where alternating layers with 8° and -8° tilting angles were stacked along [001]$_{pc,}$ [010]$_{pc}$ and [100]$_{pc}$ directions. These $a^-a^-a^-$ domains were randomly distributed and allowed to overlap. The size was less than ~10 nm across and is in good agreement with the size measured from dark-field images (less than ~8 nm across). The next step was to build the in-phase tilted domains. The ideal perovskite structure was again used as starting point followed by oxygen octahedra tilting. As enforced by in-phase tilting system ($a^0a^0c^+$) all oxygen octahedra were tilted about [001]$_{pc}$ direction. For simplicity the same value of the tilting angle was chosen for both antiphase and in-phase tilting. The $a^0a^0c^+$ in-phase tilt system is schematically represented in Fig. 6a. Three in-phase tilted domain variants with $a^0a^0c^+$, $a^0c^+a^0$ and $c^+a^0a^0$ tilt systems have been randomly embedded in the antiphase tilted matrix and allowed to overlap. Following parameters have also been thoroughly tested: antiphase/in-phase tilted system ratio, size of domains and shape of domains. The antiphase/in-phase tilt ratio is considered 100/0 when no in-phase domains are in the structure and 50/50 when 50% of the antiphase tilted matrix has been replaced by in-phase domains. Different ratios (100/0, 90/10, 70/30, 50/50, 30/70 and 10/90) have been tested as illustrated in Supplementary Fig. 3. For these simulations three in-phase domain variants of 10×10×2, 10×2×10 and 2×10×10 unit cells large were used. The short dimension of tetragonal domains corresponds to the direction about which the oxygen octahedra were tilted, namely $[001]_{pc}$, $[010]_{pc}$ and $[100]_{pc}$. The experimental data revealed the presence of diffuse scattering rods in *hk0.5* slice with increased intensity at positions of *½(ooe)* reflections. This diffuse scattering feature is the most prominent one, so in order to compare different parameters the *hk0.5* reciprocal-space plane was calculated using the kinematical approximation. Comparing the morphology and intensity of the diffuse scattering rods in the models with different antiphase/in-phase tilting ratios (Supplementary Fig. 3), one can conclude that the concentration of in-phase domains needs to be higher than 50% for a good match between the experimental and simulated data to be obtained. This indicates that the studied composition (85NBT-10BKT-5BT) is closer to the tetragonal side of the MPB rather than the rhombohedral side. This is in good agreement with the reported tetragonal average structure for this composition[10].

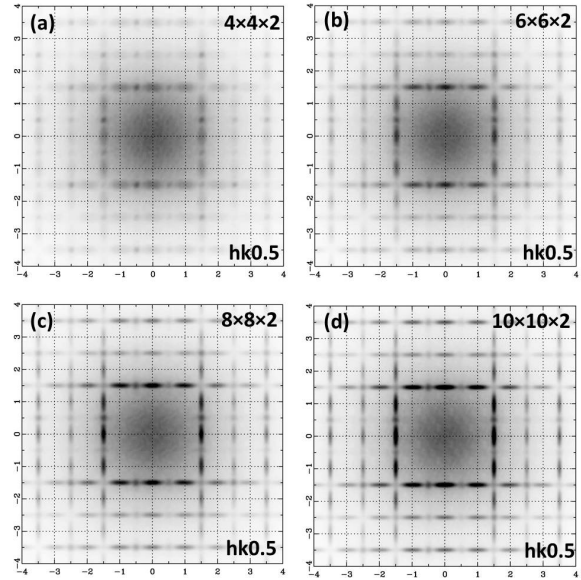

**Figure 5** | **Simulated electron diffraction patterns in the reciprocal-space plane *hk0.5* with different sizes of the in-phase domains.** All patterns have been calculated from a 50×50×50 supercell with an antiphase/in-phase ratio of 30/70 but using different initial sizes for the in-phase domains: **(a)** 4×4×2 unit cells, **(b)** 6×6×2 unit cells, **(c)** 8×8×2 unit cells and **(d)** 10×10×2 unit cells. Electron diffraction patterns have been calculated using the kinematical approximation and by averaging the results from 20 different simulations, in order to reduce the anisotropic contribution to the diffraction intensities.

Another parameter that has been tested is the size of the in-phase domains, as can be seen in Fig. 5. For these simulations the antiphase/in-phase ratio was 30/70 and the initial size of the in-phase domains varied from 4×4×2, 6×6×2, 8×8×2 to 10×10×2 unit cells large. Similar with the previous simulations three in-phase domain variants were used for each different size i.e. 4×4×2, 4×2×4 and 2×4×4. The in-phase domains were randomly distributed and allowed to overlap. The initial size of the in-phase domains represents the minimum size of an isolated in-phase domain. A good match between experimental and calculated data was obtained for sizes higher than 4×4×2 unit cells. Since the in-phase domains were allowed to overlap, large domains (less than ~14 nm across) were created and this is in good agreement with the measured size from dark-field images (less than ~12 nm across). In Fig. 6c a small region of a 2D slice cut from a 50×50×50 supercell depicting the three in-phase variants (blue, green and magenta octahedra) randomly distributed in the antiphase tilted matrix (red octahedra) is shown. Due to the fact overlapping of domains is allowed, large in-phase domains are created.

The high concentration of in-phase domains (70%) is indicated by the low numbers of antiphase tilted octahedra (red color).

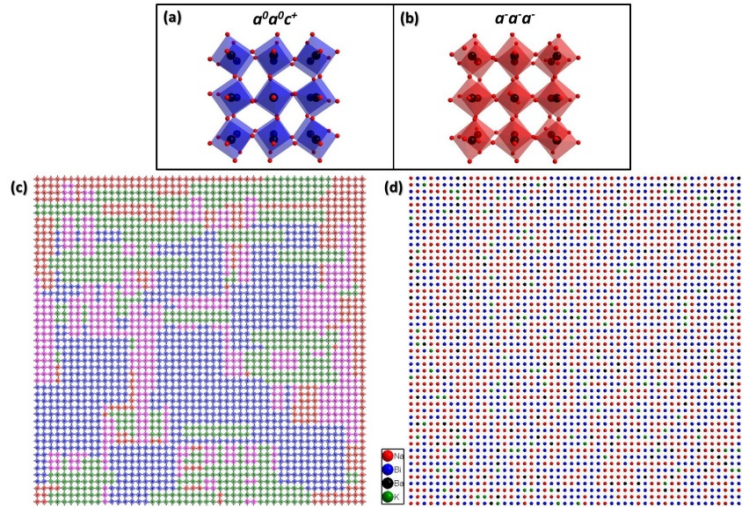

**Figure 6**. **Schematic representation of local structural disorder in 85NBT-10BKT-5BT ceramic along [001]$_{pc}$ direction. (a)** Schematic representation of in-phase $a^0a^0c^+$ tilt system. **(b)** Schematic representation of antiphase $a^-a^-a^-$ tilt system. **(c)** A small region of a 2D slice cut from a 50×50×50 supercell showing the three in-phase variants randomly distributed in the antiphase tilted matrix (red octahedra tilted according to $a^-a^-a^-$ tilt system). The blue octahedra are tilted according to $a^0a^0c^+$ tilt system while the green and magenta octahedra are tilted according to $a^0c^+a^0$ and $c^+a^0a^0$ tilt systems. The antiphase/in-phase ratio is 30/70 and the initial sizes of in-phase variants are 10×10×2, 10×2×10 and 2×10×10 unit cells large. Only the TiO$_6$ octahedra are displayed. **(d)** A 2D slice through the 50×50×50 supercell along [001]$_{pc}$ direction illustrating the Na/Bi short range chemical order. For simplicity only the A-site cations are displayed: Na(red), Bi (blue), Ba (black) and K (green).

The last tested parameter was the shape of in-phase domains. By shape is meant that the short dimension of in-phase tilted domains was varied from 2 unit cells large up till 10 unit cells large (Supplementary Fig. 4). The short coherence length of in-phase tilting (~1 nm) seems to be crucial for obtaining a good match between experimental and simulated data. The diffuse scattering rods gradually transform into sharp ½(ooe) reflections with increasing the short dimension of in-phase tilted domains from 4 up till 10 unit cells large. Analyzing all simulations one can conclude that tetragonal platelets develop along the three equivalent pseudocubic directions and are randomly distributed within a rhombohedral matrix. Up till now this model only explains the diffuse scattering rods observed in *hk0.5*-slice of the RED data. To further check our model electron diffraction patterns were simulated along $[100]_{pc}$, $[011]_{pc}$ and $[111]_{pc}$ zone axes (Supplementary Fig. 1(d-f)). When comparing to the experimental data a good agreement is obtained within the limits of the kinematical approximation. All superstructure reflections (½*(ooe)* and ½*(ooo)*) are accounted for with a model which has an antiphase/in-phase tilting ratio of 30/70 and in-phase tilted domains of $10\times10\times2$ unit cells large. To this point the model explains two out of the three features observed experimentally, namely the diffuse scattering rods and superstructure reflections. In order to account for the broad diffuse intensity near the fundamental perovskite reflections Na/Bi occupancy disorder and local cation displacements correlated along both $<111>_{pc}$ and $<001>_{pc}$ directions were included in the model. The degree of chemical short-range order (CSRO) is in fact very small but is significantly different from a random distribution. To obtain the CSRO a positive correlation of 0.1 was used in the $<001>_{pc}$ directions together with a negative correlation of -0.1 in the $<111>_{pc}$ directions. Ba and K atoms are randomly introduced in the structure which leads to a suppression of Na/Bi A-site CSRO. A 2D slice cut from the $50\times50\times50$ supercell illustrating the Na/Bi occupancy disorder is shown in Fig. 6d. A small tendency of alternating rows with Na- and Bi-containing unit cells along $<001>_{pc}$ directions can be observed. Furthermore Bi-Bi distances along $<111>_{pc}$ and $<001>_{pc}$ directions have been increased by 1-2% while Na-Na distances have been decreased by 1-2%. Ba, K and Ti atoms were assumed to remain at their average positions. Figure 7 illustrates the simulated electron diffraction patterns in the *hk0* and *hk0.5* reciprocal-space planes calculated from the final model. Taking into consideration that the kinematical approximation was used for the simulations, a good match was obtained between the experimental patterns and the simulated ones as can be seen in Fig. 7(c-h). Both the morphology and intensity distribution (i.e. higher intensity at positions of ½*(ooe)* reflections) are accounted for by the model, in the case of the diffuse scattering rods present in *hk0.5*-slice. However, in the case of the broad diffuse scattering intensity near the fundamental perovskite reflections (Fig. 7c,d), which is observed around the *hk0* plane, the calculated diffuse scattering intensity (Fig. 7e,f) deviates slightly from the experimental one. This might be due to the fact that the broad diffuse intensity near the fundamental perovskite reflections is affected more by dynamical effects and absorption. Moreover, the simulated patterns were calculated using the kinematical approximation and only one wavelength of 0.025 Å. In reality the electrons in the microscope present an energy spread (~0.9 eV) which was not taken into consideration and might affect the very weak diffuse scattering intensity.

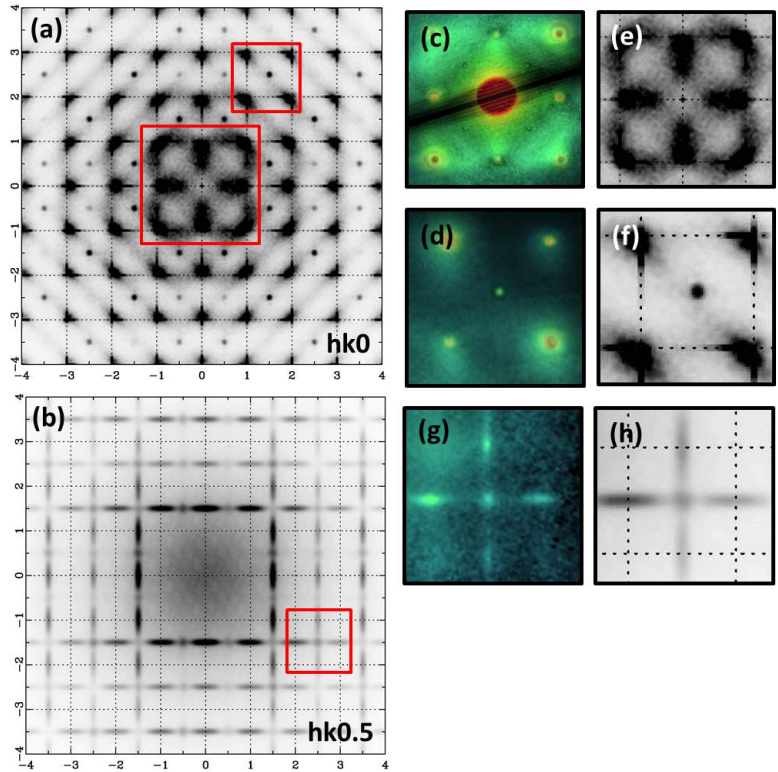

**Figure 7. Simulated electron diffraction patterns. (a)** *hk0* reciprocal-space plane. **(b)** *hk0.5* reciprocal-space plane. **(c), (d)** Enlarged images of the regions depicted by white squares in the experimental *hk0* pattern (Fig. 3b) compared with the corresponding calculated region **(e), (f)**. **(g)** Enlarged image of the region depicted by the white square in the experimental *hk0.5* pattern (Fig. 3c) compared with the corresponding calculated region **(h)**. The diffraction patterns were calculated using the kinematical approximation from a 50×50×50 supercell with an antiphase/in-phase ratio of 30/70 and in-phase tilted domains of 10×10×2 unit cells large. A small degree of Na/Bi CSRO together with Na/Bi correlated displacements along $<001>_{pc}$ and $<111>_{pc}$ directions were also considered. The simulated electron diffraction patterns were obtained by averaging the results from 20 different simulations, in order to reduce the anisotropic contribution to the diffraction intensities.

The final model consists of three variants of in-phase domains tilted about the three pseudocubic axes randomly distributed into an antiphase tilted matrix ($a^-a^-a^-$). In addition Na/Bi CSRO was introduced on the A-site together with correlated Na- and Bi-displacements along $<111>_{pc}$ and $<001>_{pc}$ directions. The two types of octahedral tilting are responsible for the two types of superstructure reflections while the plate-like shape of in-phase domains accounts for the diffuse scattering rods present in *hk0.5*-type slices. The broad diffuse intensity near the fundamental perovskite reflections can be explained by a small degree of CSRO on the A-site in combination with correlated Na- and Bi-displacements.

## Discussion

So far, many structural investigations have reported local deviations from the long-range structure for NBT[26,33] and NBT-based solid solutions[29-31] but the nature of these local deviations is still a matter of controversy. In this work a structural model of the local disorder in 85NBT-10BKT-5BT ceramic was developed by analyzing the intensity and morphology of Bragg reflections and diffuse scattering features in 3D. The RED method[42,43] proved to be well suited for simultaneously recording sharp Bragg reflections and weak diffuse scattering intensity, uncovering new information in 3D. Three extra features besides the fundamental perovskite reflections were revealed by the RED data: (i) two types of superstructure reflections ½($ooe$) and ½($ooo$), (ii) broad diffuse scattering intensity around the fundamental perovskite reflections and (iii) continuous diffuse scattering rods present in $hk0.5$-type slices. The simultaneous analysis of intensity and morphology of both superstructure reflections and diffuse scattering intensity along different reciprocal-space planes facilitated the development of a complex structural model. The model we developed explains all observed features and consists of plate-like tetragonal domains (in-phase tilting) developed along the three equivalent pseudocubic directions which are randomly distributed within a rhombohedral matrix (antiphase tilting). In addition Na/Bi occupancy disorder on the A-site and correlated Na- and Bi-displacements along <111>$_{pc}$ and <001>$_{pc}$ directions have also been introduced in the final model. Our model suggests that the studied composition is situated on the tetragonal side of the MPB rather than the rhombohedral one with an approximate antiphase/in-phase tilt ratio of 30/70. The high concentration of in-phase tilted domains is in good agreement with the reported tetragonal average structure[10]. The size of the antiphase domains was less than ~10 nm across while for in-phase domains the size was less than ~14 nm across. Similar to other NBT-compounds[54,55] the octahedral tilting for the studied composition is confined to the nanoscale which makes it difficult to assign a long-range octahedral tilt system. Several previous studies[32-34] have already reported the presence of nanosized tetragonal plate-like domains embedded in a rhombohedral matrix for pure NBT, but this is the first time when a comprehensive structural model is developed that explains all observed diffraction features. In the case of pure NBT these tetragonal plate-like domains are formed at high temperature and are associated with a residual tetragonal phase within the rhombohedral matrix. In our case the addition of Ba and K stabilizes the tetragonal domains which grow and are larger in size than the rhombohedral ones. Moreover the concentration of tetragonal domains is increased suggesting that the studied composition is closer to the tetragonal side of the MPB. On the other hand bond valence calculations[14] showed that $Bi^{3+}$ atoms are significantly underbonded while $Na^{1+}$ atoms are slightly overbonded indicating a highly distorted local environment, especially for $Bi^{3+}$ atoms. The requirement for $Bi^{3+}$ to improve its coordination environment is the main drive for the local displacements along $<111>_{pc}$ and $<001>_{pc}$ directions. Similar behavior was also observed for Pb-based ferroelectrics in PZT and PMN[61,62]. CSRO on the A-site has also been reported for pure NBT and NBT-BT and was attributed to a tendency of reducing charge imbalance in the bulk structure[26,29,31]. The very small degree of Na/Bi chemical short range order might be the reason why it had eluded direct experimental evidence. Our model is also in good agreement with the GPZ-model[26] for pure NBT which implies a low degree of A-site CRSO and correlated Na- and Bi-displacements along both $<111>_{pc}$ and $<001>_{pc}$ directions. Since the GPZs are planar defects which give rise to diffuse scattering rods in reciprocal-space they could be analogous to the plate-like in-phase domains from our model. The current study reveals the complexity of the local structure of 85NBT-10BKT-5BT ternary ceramic.

**Methods**

**TEM characterization.** The RED data and dark-field images were recorded at room temperature using a JEOL JEM-2100F microscope with ultra-high resolution pole-piece operated at 200 kV. For both experiments a Gatan Orius 200D CCD camera was used to record the electron diffraction patterns and dark-field images, respectively. TEM specimens were prepared via mechanical polishing till a thickness of about 30 μm, followed by Ar ion milling (Fishicone Model 1050).

**RED method.** An aperture with a diameter of 500 nm was used to select the region of interest during RED data collection. The data collection is semi-automated within the REDaquisition[43] software. This method combines the mechanical tilt of the goniometer with a fine beam tilt which allows a very fine sampling of reciprocal-space. The goniometer tilting range was ±21° with a tilt step of 2° while the beam tilting range was ±1° with a beam tilt step of 0.1°. In total 440 electron diffraction frames were collected with 1.2 s acquisition time for each frame. Moreover, the collection of RED data does not require any alignment along exact zone axis patterns which means that diffuse scattering intensity is recorded in the presence of fewer Bragg reflections excited simultaneously. The reconstruction of 3D reciprocal-space volume was done using REDprocessing[43] software while for the 3D visualization VolView software was used.

**Simulations.** The structural model for local disorder in 85NBT-10BKT-5BT ternary ceramic was developed using DISCUS[47,48] software. In order to randomly distribute the origins of antiphase/in-phase domains an intrinsic function of the software was used, which returns Gaussian distributed pseudo random numbers with mean zero. To create chemical short-range order Monte Carlo simulations were performed. Within DISCUS software the occupancy of a site in the crystal is described using pseudo Ising spin variables: $\sigma_i = +1$ or $\sigma_i = -1$. If $\sigma_i = 1$ it means site $i$ is occupied by atom $A$ while if $\sigma_i = -1$ then site $i$ is occupied by atom $B$. Using this variables the energy, $E_{occ}$, takes the following form:

$$E_{occ} = \sum_i H\sigma_i + \sum_i \sum_{n,n\neq i} J_n \sigma_i \sigma_{i-n} \quad (2)$$

The sums are over all sites $i$ and neighbors $n$ in the supercell. The value $\sigma_{i-n}$ refers to the occupancy (spin) of the neighboring site $i-n$ of site $i$. $J_n$ are pair interaction energies corresponding to the neighboring vector defined by $i$ and $n$. The quantity $H$ is the energy of a single site and controls the overall concentration. If two neighboring sites are occupied by the same atom type the product $\sigma_i \cdot \sigma_{i-n} = 1$ and if the energy term $J_n$ is positive alike pairs will lead to a larger energy and thus be avoided. On the contrary if $J_n$ is negative alike pairs will be favored. In the case of positive correlations two neighboring sites are occupied by the same atom type while a negative correlation will cause the two sites to be preferably occupied by different atom types. Na- and Bi-displacements were created also via Monte Carlo simulations using instead a Lennard-Jones potential, as in the following equation:

$$E_{ij} = \sum_i \sum_{n,n\neq i} D\left[\left(\frac{\tau_{in}}{d_{in}}\right)^{12} - 2\left(\frac{\tau_{in}}{d_{in}}\right)^6\right] \quad (3)$$

The sums are over all atom sites $i$ in the supercell and all neighboring sites $n$. $D$ represents the potential depth which must be negative, $\tau_{in}$ is the target distance where Lennard-Jones potential has its minimum and $d_{in}$ represents the user chosen distance between two neighboring atoms. The Monte Carlo algorithm is used to minimize the energy of a model system (i.e. our supercell) in order to create either CSRO or displacive disorder. The different energy terms for CSRO and displacive disorder are presented in equations (2) and (3), respectively. In the case of CSRO two different atoms are selected at random and their places are switched. In our case the constraints were a positive correlation of 0.1 in the <001>$_{pc}$ directions and a negative correlation of -0.1 in the <111>$_{pc}$ directions. For displacive disorder a randomly selected atom is shifted by a random amount along the user defined directions. In our case the displacements were limited to <111>$_{pc}$ and <001>$_{pc}$ directions. The energy difference of the configuration before and after a change is computed. The move is accepted if the energy difference is less than zero or if the probability $P$, given by eqn. (4), is less than a random number chosen uniformly in the range $[0,1]$.

$$P = \frac{e^{-\frac{\Delta E}{kT}}}{1 + e^{-\frac{\Delta E}{kT}}} \quad (4)$$

Where $T$ is the temperature and $k$ the Boltzmann's constant. This process is repeated until the energy reaches a minimum.

**Acknowledgement**

The Knut and Alice Wallenberg (KAW) Foundation is acknowledged for funding for the electron microscopy facilities at Stockholm University and financial support for A.N. and C.W.T. under the project 3DEM- NATUR.

**Author contributions**

C.-W.T. conceived the idea of the study, designed the experiments, and supervised the project. A.N. designed and performed all experiments and computer simulations, and wrote the manuscript. Both authors contributed to data analysis and finalized the manuscript.

# Supplementary information

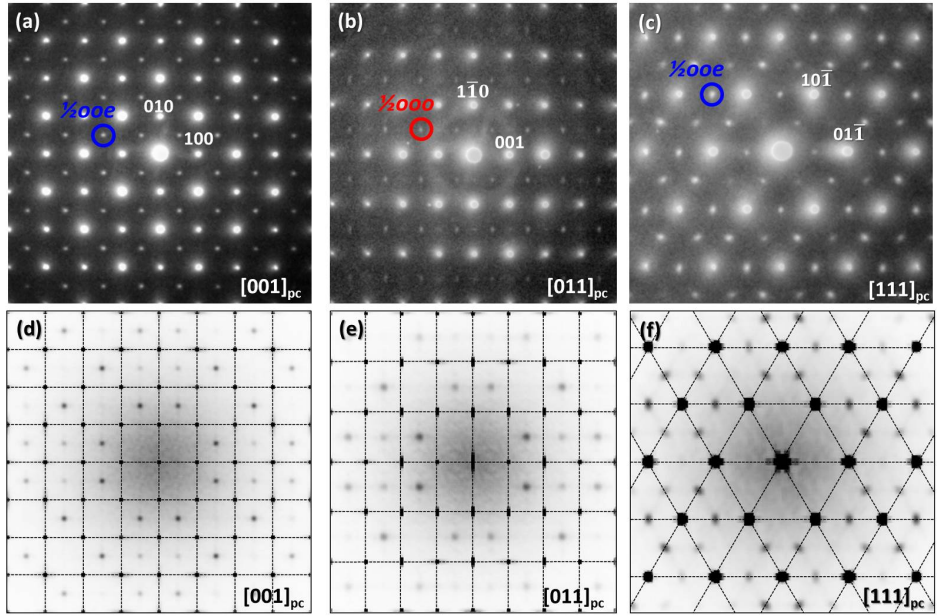

**Supplementary Figure 1. Experimental SAED patterns and kinematical simulated electron diffraction patterns for 85NBT-10BKT-5BT ternary compound.** The experimental SAED patterns have been recorded along **a.** $[001]_{pc}$, **b.** $[011]_{pc}$ and **c.** $[111]_{pc}$ zone axes. The simulated electron diffraction patterns were calculated along the same zone axes, namely **d.** $[001]_{pc}$, **e.** $[011]_{pc}$ and **f.** $[111]_{pc}$. For simplicity and convenience of comparison pseudocubic axes ($Pm\bar{3}m$) have been used for indexing.

Two types of superstructure reflections with respect to the ideal cubic perovskite structure can be observed, namely ½($ooe$) as depicted by blue circles in ZAs $[001]_{pc}$ and $[111]_{pc}$ and ½($ooo$) as depicted by the red circle in ZA $[011]_{pc}$. All patterns have been calculated from a 50×50×50 supercell with initial in-phase tilted domains of 10×10×2 unit cells large and a 30/70 antiphase/in-phase tilting ratio. At this stage no cation displacements or short range chemical order has been considered. The kinematical approximation was used for calculating the electron diffraction patterns. In the case of experimental SAED patterns, kinematically forbidden reflections often appear due to the occurrence of double diffraction. For example the $a^0a^0c^+$ tilt system allows for superstructure reflections that have $h \neq \pm k$. Hence, the *1/2 1/2 0* reflection in ZA $[001]_{pc}$ is kinematically forbidden but appears due to double diffraction, such as the route *3/2 3/2 0* ± 1-10. In the case of $a^-a^-a^-$ tilt system the condition for allowed superstructure reflections is $h \neq \pm k$, $k \neq \pm l$ and $l \neq \pm h$ and similar to the *1/2 1/2 0* reflection the *1/2 1/2 1/2* reflection in ZA [011]$_{pc}$ is also forbidden and appears due to double diffraction. Taking into consideration double diffraction and the fact that the kinematical approximation was used for calculating the simulated electron diffraction patterns, a good agreement was obtained between the experimental patterns and the simulated ones. All superstructure reflections have been accounted for with a two-tilt system model which includes the $a^0a^0c^+$ and $a^-a^-a^-$ tilt systems.

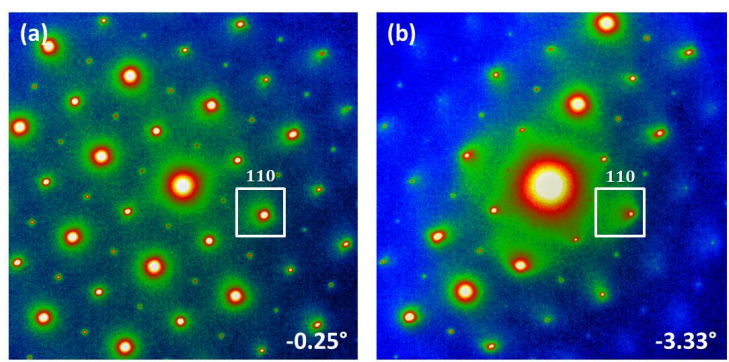

**Supplementary Figure 2. Experimental SAED patterns. a.** Electron diffraction pattern very close to $[001]_{pc}$ ZA. **b.** Electron diffraction pattern ~3° away from $[001]_{pc}$ ZA. Broad diffuse scattering intensity near the fundamental perovskite reflections can be clearly observed in (b) as highlighted by the white square for 110 reflection. Howeve,r this is not the case for the electron diffraction pattern (a) which is very close to exact ZA condition.

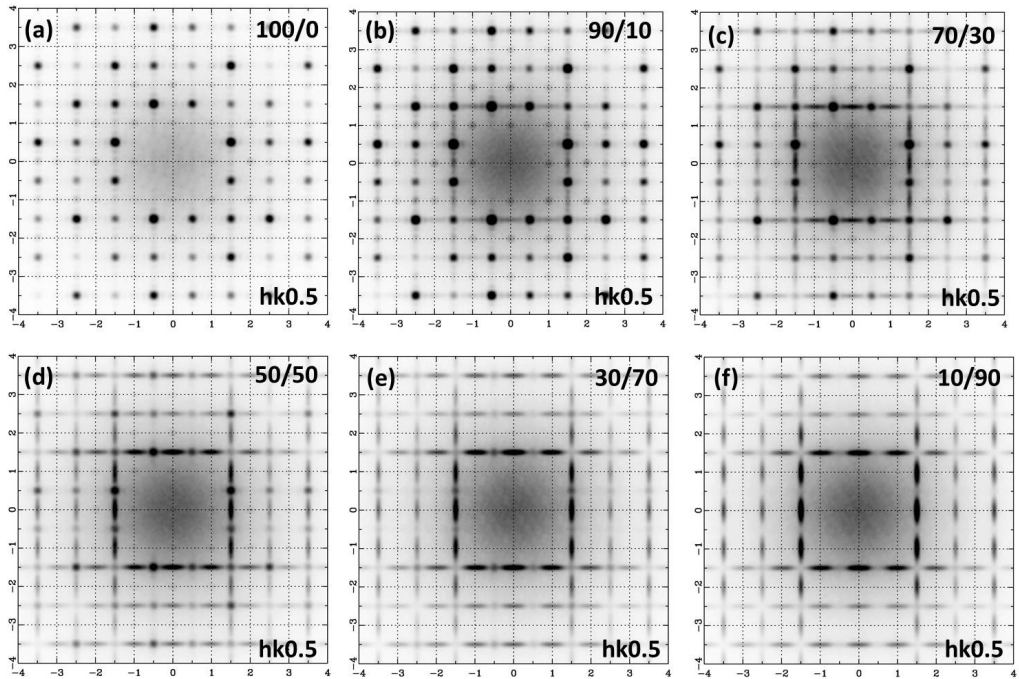

**Supplementary Figure 3. Simulated electron diffraction patterns in the reciprocal-space plane *hk0.5* with different antiphase/in-phase tilting ratios.** All patterns have been calculated from a 50×50×50 supercell with tetragonal domains of 10×10×2 unit cells large but with different antiphase/in-phase ratios: **a.** 100/0, **b.** 90/10, **c.** 70/30, **d.** 50/50, **e.** 30/70 and **f.** 10/90. Electron diffraction patterns have been calculated using the kinematical approximation and by averaging the results from 20 different simulations, in order to reduce the anisotropic contribution to the diffraction intensities.

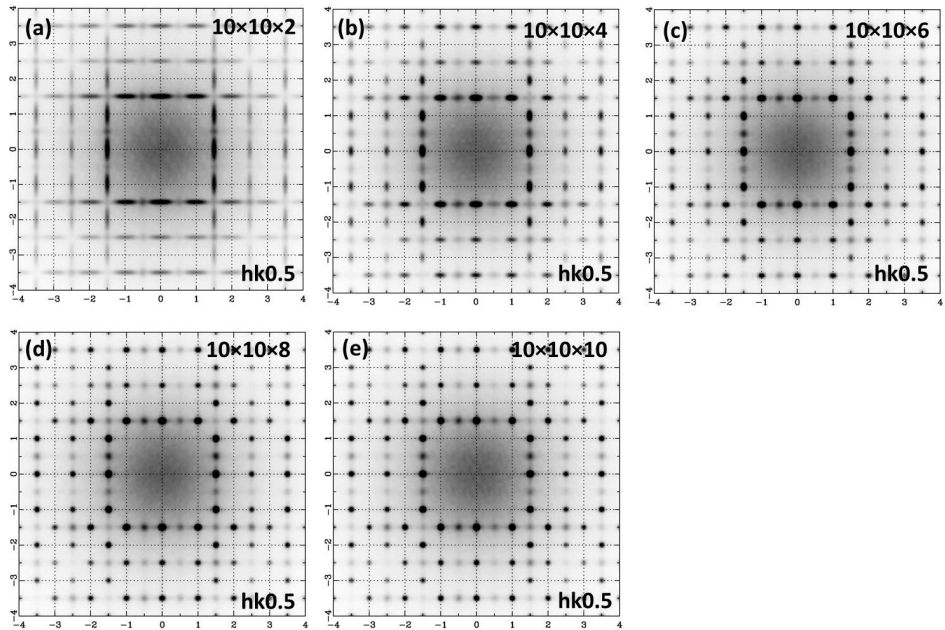

**Supplementary Figure 4. Simulated electron diffraction in the reciprocal-space plane *hk0.5* with different shapes for the in-phase domains.** All patterns have been calculated from a 50×50×50 supercell with an antiphase/in-phase ratio of 30/70 but using different shapes for the in-phase domains: **a.** 10×10×2 unit cells **b.** 10×10×4 unit cells **c.** 10×10×6 unit cells, **d.** 10×10×8 and **e.** 10×10×10 unit cells. Electron diffraction patterns have been calculated using the kinematical approximation and by averaging the results from 20 different simulations.